\documentclass[prc,twocolumn,floatfix,groupedaddress,nofootinbib,showpacs,preprintnumbers,
amsmath,amssymb,amsfonts,superscriptaddress,widetable] {revtex4}
\usepackage{bm}
\usepackage{mathrsfs}
\usepackage{amssymb}
\usepackage{amsmath}
\usepackage{graphicx}
\usepackage{array}
\usepackage{xcolor}

\begin{document}
\title{GW190814: Impact of a 2.6 solar mass neutron star \\ on nucleonic equations of state}
\author{F. J. Fattoyev}\email{ffattoyev01@manhattan.edu}
\affiliation{Department of Physics, Manhattan College,
                Riverdale, NY 10471, USA}
\author{C. J. Horowitz}\email{horowit@indiana.edu}
\affiliation{Center for Exploration of Energy and Matter and
                  Department of Physics, Indiana University,
                  Bloomington, IN 47405, USA}
\author{J. Piekarewicz}\email{jpiekarewicz@fsu.edu}
\affiliation{Department of Physics, Florida State University,
               Tallahassee, FL 32306, USA}
\author{Brendan Reed}\email{reedbr@iu.edu}
\affiliation{Department of Astronomy, Indiana University, 
                Bloomington, Indiana 47405, USA}

\date{\today}
\begin{abstract}
Is the secondary component of GW190814 the lightest black hole or the heaviest neutron star ever discovered in a double compact-object system [R. Abbott {\sl et al.} ApJ Lett., 896, L44 (2020)]? This is the central question animating this letter. Covariant density functional theory provides a unique framework to investigate both the properties of finite nuclei and neutron stars, while enforcing causality at all densities. By tuning existing energy density functionals we were able to: (a) account for a $2.6\,M_{\odot}$ neutron star, (b) satisfy the original constraint on the tidal deformability of a $1.4\,M_{\odot}$ neutron star, and (c) reproduce ground-state properties of finite nuclei. Yet, for the class of models explored in this work, we find that the stiffening of the equation of state required to support super-massive neutron stars is inconsistent with either constraints obtained from energetic heavy-ion collisions or from the low deformability of medium-mass stars.
Thus, we speculate that the maximum neutron star mass can not be significantly higher than the existing observational limit and that the $2.6\,M_{\odot}$ compact object is likely to be the lightest black hole ever discovered. 
\end{abstract}
\smallskip
\pacs{
21.60.Jz,   
24.10.Jv,   
26.60.Kp,   
97.60.Jd   
}

\maketitle

The first direct detection of gravitational waves from the binary collision of two black holes launched the new era of gravitational-wave astronomy\,\cite{Abbott:PRL2016}.Two years later, the detection of gravitational waves from GW170817---a binary neutron star merger\,\cite{Abbott:PRL2017}---in  association with its electromagnetic counterpart\,\cite{Drout:2017ijr,Cowperthwaite:2017dyu, Chornock:2017sdf,Nicholl:2017ahq}, greatly advanced multimessenger astronomy. And two years after GW170817, the LIGO-Virgo collaboration continues to mesmerize the physics community after reporting the detection of gravitational waves from the coalescence of a binary system with the most extreme mass ratio ever observed: a 23 solar mass black hole and a 2.6 solar mass ``compact" object\,\cite{Abbott:2020khf}.  Although data from all three instruments (LIGO-Livingston, LIGO-Hanford, and Virgo) allowed good sky localization of the source, no electromagnetic counterpart has been reported. Moreover, unlike GW170817, no measurable tidal signature was imprinted on the gravitational waveform, which seems consistent with the relatively large mass of the 2.6\,$M_{\odot}$ compact object. Hence, one is left speculating whether the compact object is either the most massive neutron star or the lightest black hole ever discovered\,\cite{Most:2020bba,Tan:2020ics}.

The discovery paper suggests that, based on several current estimates of the maximum neutron star mass, ``GW190814 is unlikely to originate in a neutron star-black hole (NSBH) coalescence"\,\cite{Abbott:2020khf}. Yet the paper leaves open the possibility that improved knowledge of the equation of state or further observations could alter this assessment. The absence of super-massive neutron stars is consistent with the analysis by Margalit and Metzger who argue against their formation based on the lack of evidence of a large amount of rotational energy in the ejecta during the spin-down phase of GW170817\,\cite{Margalit:2017dij}. Interestingly, the suggested upper limit of $M_{\rm max}\!\lesssim\!2.17\,M_{\odot}$\,\cite{Margalit:2017dij} is in full agreement with the recent observation by Cromartie and collaborators of the heaviest $2.14^{+0.10}_{-0.09}\,M_{\odot}$ neutron star to date\,\cite{Cromartie:2019kug}. However, given that one can not definitively exclude the existence of super-massive neutron stars, we explore here the impact of a 2.6\,$M_{\odot}$ neutron star on nucleonic equations of state, particularly in the framework of covariant density functional theory. 

It has been known for more than two decades that the class of covariant energy density functionals (EDFs) used in this work can reproduce nuclear observables at normal nuclear densities and also generate neutron stars with maximum masses that differ by more than one solar mass\,\cite{Mueller:1996pm}. Hence, stable neutron stars with 2.6\,$M_{\odot}$---and even higher---can be readily generated. The challenge, however, is not to reconcile super-massive neutron stars with the properties of finite nuclei, but rather, with neutron-star properties that are sensitive to the equation of state (EOS) at two-to-three times nuclear densities, such as stellar radii and tidal deformabilities that favor a rather soft EOS\,\cite{Abbott:PRL2017,Abbott:2018exr}.

In particular, GW170817 has provided stringent constraints on the EOS of neutron rich matter at a few times nuclear densities from the determination of the tidal deformability of a $M\!=\!1.4\,M_{\odot}$ neutron star\,\cite{Bauswein:2017vtn, Fattoyev:2017jql, Annala:2017llu, Abbott:2018exr, Most:2018hfd, Tews:2018chv, Malik:2018zcf, Tsang:2018kqj, Radice:2017lry, Radice:2018ozg, Tews:2019cap, Capano:2019eae, Drischler:2020yad, Xie:2020tdo, Chatziioannou:2020pqz}. Constraints from GW170817 seem to favor compact stars with relatively small stellar radii, suggesting a relatively soft EOS. These constraints are consistent with the recent determination of both the mass and radius of PSR J0030+0451 by the Neutron star Interior Composition Explorer (NICER)\,\cite{Riley:2019yda,Miller:2019cac}. Pulse-profile modeling of the thermal emission from the pulsar's hot spots suggest a mass of about 1.4\,$M_{\odot}$ and a radius of nearly 13\,km, with a $\pm\,10\%$ uncertainty on both quantities\,\cite{Riley:2019yda,Miller:2019cac}. Note that although consistent with GW170817, NICER results can accommodate slightly stiffer equations of state. However, a real tension develops as one aims to reconcile a relatively soft EOS as demanded by GW170817, with the much stiffer EOS required to account for heavy neutron stars with masses in the vicinity of 2$M_{\odot}$\,\cite{Demorest:2010bx, Antoniadis:2013pzd, Cromartie:2019kug}. Based on this combined evidence, suggestions have been made for the existence quark matter cores in massive neutron stars\,\cite{Annala:2019puf}. Although the claim may be premature, after all there are purely nucleonic EOS that satisfy all experimental and observational constraints to date\,\cite{Fattoyev:2017jql}, the prospect of identifying an assumed phase transition in the stellar cores is exciting. As shown below, the tension is exacerbated if the EOS must be stiffened even further to account for the possible existence of super-massive neutron stars. 

To test the possible existence of a 2.6\,$M_{\odot}$ neutron star we rely on density functional theory\,\cite{Kohn:1999}.
The energy density functional employed here is defined in terms of an underlying Lagrangian density that has been extensively discussed in earlier publications\,\cite{Horowitz:2000xj, Todd-Rutel:2005fa, Chen:2014sca}, so we limit ourselves to highlight those terms of relevance to the high-density component of the equation of state. That is,
\begin{equation}
{\mathscr L}\!=\!\ldots +
    \frac{\zeta}{4!}  g_{\rm v}^4(V_{\mu}V^\mu)^2 +
   \Lambda_{\rm v}\Big(g_{\rho}^{2}\,{\bf b}_{\mu}\cdot{\bf b}^{\mu}\Big)\!
                           \Big(g_{\rm v}^{2}V_{\nu}V^{\nu}\Big).
 \label{LDensity}
\end{equation}
The basic degrees of freedom of the model are neutrons and protons interacting via the exchange of photons and ``mesons".  Besides the conventional Yukawa couplings (not shown) the model includes non-linear meson interactions that serve to simulate the complicated many-body dynamics and that are required to improve the predictive power of the model. The two terms shown in Eq.~(\ref{LDensity}) fall into this category and are of critical importance to the behavior of dense, neutron-rich matter. The first term in the expression describes a quartic self-interaction of the isoscalar-vector field $V^{\mu}$ 
which affects the EOS of symmetric nuclear matter at high densities\,\cite{Mueller:1996pm}.
In turn, the last term includes a mixed quartic coupling between $V^{\mu}$ and the isovector-vector field ${\bf b}^{\mu}$. This term was introduced to modify the density dependence of the symmetry energy, which plays a critical role in the structure of both neutron-rich nuclei and neutron stars\,\cite{Horowitz:2000xj}. Note that the symmetry energy is to a very good approximation equal to the difference in the energy per nucleon between pure neutron matter and symmetric nuclear matter. 
Covariant density functional theory provides a relativistic consistent framework as one extrapolates to dense matter as it
ensures---unlike non-relativistic formulations---that the EOS remains causal at all densities. 
Finally, the structure of neutron stars will be explored by enforcing both charge neutrality and chemical equilibrium. As such, the basic constituents of the model are nucleons and leptons (both electrons and muons). No ``exotic" degrees of freedom---such as hyperons, meson condensates, or quarks---will be considered.

As already alluded, tuning the $\zeta$ parameter in Eq.~(\ref{LDensity}) allows one to stiffen the symmetric-matter EOS to produce super-massive neutron stars. For example, two of the EDFs used in Ref.\,\cite{Fattoyev:2017jql}---IUFSU\,\cite{Fattoyev:2010mx} and FSUGarnet\,\cite{Chen:2014mza})---that are consistent with both the 2$M_{\odot}$ constraint\,\cite{Demorest:2010bx, Antoniadis:2013pzd, Cromartie:2019kug} and the tidal deformability of a 1.4\,$M_{\odot}$ neutron star\,\cite{Abbott:PRL2017}, can be adjusted to produce maximum stellar masses of at least 2.8\,$M_{\odot}$. However, we find no need to strain the model to such an extreme, so we tune $\zeta$ to produce a maximum neutron star mass of 2.6\,$M_{\odot}$. 
Reducing the $\zeta$ parameter requires tuning the other model parameters. To do so, we start with the bulk properties 
predicted by IUFSU---together with the analytical transformation described in\,\cite{Chen:2014sca}---to 
connect the bulk properties to the model parameters. Next, we slightly decrease the value of the slope of the symmetry 
energy to obtain a neutron skin of $0.15$\,fm for $^{208}$Pb---and adjusted the value of the incompressibility coefficient 
to $K\!=\!227$\,MeV to maintain agreement with the centroid energy of the giant monopole resonance in 
$^{208}$Pb.
Finally, 
we re-adjust the values of $m_{\rm s}$, the saturation density $\rho_0$, and the binding energy at saturation 
$\epsilon_0$ to ensure that the binding energies and charge radii of both $^{40}$Ca and $^{208}$Pb remain intact.
We refer to this nuclear EDF as ``BigApple" and display its parameters, as defined in Ref.\,\cite{Chen:2014sca}, in Table\,\ref{Table0} alongside the other two covariant EDFs used in this work.

\begin{widetext}
\begin{center}
\begin{table}[h]
\begin{tabular}{|l||c|c|c|c|c|c|c|c|}
 \hline\rule{0pt}{2.5ex} 
 Model & $m_{\rm s}$  
       & $g_{\rm s}^2$ & $g_{\rm v}^2$ & $g_{\rho}^2$
       & $\kappa$ & $\lambda$ & $\zeta$ & $\Lambda_{\rm v}$\\
 \hline
 \hline
 IUFSU    & 491.500 &  99.4266 & 169.8349 & 184.6877
               & 3.38081 & $+$0.000296 & 0.03000 & 0.046000 \\
FSUGarnet & 496.939 & 110.3492 & 187.6947 & 192.9274
               & 3.26018 & $-$0.003551 & 0.02350 & 0.043377 \\
BigApple  & 492.730 & 93.5074 & 151.6839 & 200.5562
               & 5.20326 & $-$0.021739 & 0.00070 & 0.047471 \\
\hline
\end{tabular}
\caption{Parameters, as in Ref.\,\cite{Chen:2014sca}, for the three models discussed in the text. The scalar mass $m_{\rm s}$ and $\kappa$ are  in MeV. The vector-meson masses are fixed at $m_{\rm v}\!=\!782.5\,{\rm MeV}$ and $m_{\rho}\!=\!763.0\,{\rm MeV}$, respectively, and the nucleon mass at $M\!=\!939$\,{\rm MeV}.}
\label{Table0}
\end{table}
\end{center}
\end{widetext}

\begin{center}
\begin{table}[h]
\begin{tabular}{|c||c|c|c|}
\hline\rule{0pt}{2.5ex} 
\!\!Nucleus  &   $B\!/\!A(\rm MeV)$  &  $R_{\rm ch}(\rm fm)$  &  $R_{\rm skin}(\rm fm)$  \\  
 \hline
 \hline\rule{0pt}{2.5ex} 
 \!\!${}^{40}$Ca   &  8.552\,(8.551)  &   3.452\,(3.478)   &   $-0.050$   \\
 ${}^{48}$Ca       &  8.536\,(8.666)  &   3.476\,(3.477)   &   $\phantom{-}0.168$  \\ 
 ${}^{68}$Ni       &  8.643\,(8.682)  &   3.875\,(3.887)  &   $\phantom{-}0.170$   \\ 
 ${}^{90}$Zr       &  8.666\,(8.710)  &   4.255\,(4.269)  &   $\phantom{-}0.061$   \\  
 ${}^{132}$Sn      &  8.294\,(8.355)  &   4.708\,(4.709)  &   $\phantom{-}0.212$   \\ 
 ${}^{208}$Pb      &  7.868\,(7.867)  &   5.503\,(5.501)  &   $\phantom{-}0.151\,\big(0.33^{+0.16}_{-0.18}\big)$   \\ 
\hline
\end{tabular}
\caption{Theoretical predictions alongside experimental data (in parentheses) for the binding energy per nucleon\,\cite{Wang:2012} and charge radii\,\cite{Angeli:2013} for a representative set of doubly-magic and semi-magic nuclei; the experimental charge radius of ${}^{68}$Ni is from Ref.\,\cite{Kaufmann:2020gbf}. The last column displays predictions for the neutron skin thickness. With the exception of ${}^{208}$Pb\,\cite{Abrahamyan:2012gp}, no electroweak measurements of neutron skins are presently available.}
\label{Table1}
\end{table}
\end{center}
One of the central tenets of nuclear density functional theory is to provide a ``universal" EDF that can reproduce nuclear observables over an enormous range of densities and isospin asymmetries. The goal is to build a nuclear EDF that can be used to explore both the properties of finite nuclei as well as the structure of neutron stars---dynamical objects that differ in length scales by more that 18 orders of magnitude. In this context, we display in Table\,\ref{Table1} binding energies and charge radii as predicted by the BigApple.
Although the agreement between theory and experiment is fairly good, it is certainly not as good as some modern nuclear EDFs that have been fitted to a host of nuclear observables. 

Before moving on to discuss neutron star properties, we display in Table\,\ref{Table2} some bulk properties of infinite neutron-rich matter that encode its behavior in the vicinity of the saturation density; the definitions are as in Ref.\,\cite{Piekarewicz:2008nh}. The saturation point of symmetric nuclear matter is defined in terms of the saturation density $\rho_{0}$ and the value of the energy per particle $\epsilon_{0}$. Because the pressure at saturation density vanishes, the rate at which the energy per particle increases is controlled by the incompressibility coefficient $K$, a quantity that is strongly correlated to the centroid energy of the giant monopole resonance\,\cite{Harakeh:2001}. In turn, $J$ and $L$ are fundamental parameters of the symmetry energy that encode the increase in the energy per particle and its density dependence as the system becomes neutron rich. In particular, given that the pressure of symmetric nuclear matter vanishes at saturation, $L$ is closely related to the pressure of pure neutron matter at saturation density. As such, it is strongly correlated to both the neutron skin thickness of heavy nuclei and the radius of neutron stars\,\cite{Horowitz:2001ya}.
\begin{center}
\begin{table}[h]
\begin{tabular}{| l | | c | c | c || c | c |}
 \hline\rule{0pt}{2.5ex} 
\!\!Model  &  $\rho_{0}$ &  $\epsilon_{0}$ & $K$ &  $J$ & $L$ \\  
\hline
 \hline\rule{0pt}{2.5ex} 
\!\!IUFSU       & 0.155 &   $-$16.397   &   231.333    &   31.296   &  47.205 \\
FSUGarnet       & 0.153 &   $-$16.231   &   229.628    &   30.917   &  50.961  \\
BigApple        & 0.155 &   $-$16.344   &   227.001    &   31.315   &  39.800  \\
\hline
\end{tabular}
\caption{Bulk properties of nuclear matter as predicted by three covariant EDFs: IUFSU\,\cite{Fattoyev:2010mx}, FSUGarnet\,\cite{Chen:2014mza}, and BigApple. The listed properties are the saturation density of symmetric nuclear matter together with the energy per particle and incompressibility coefficient at saturation. Also shown is the value of the symmetry energy $J$ and its slope $L$ at saturation density. All quantities are given in MeV except for $\rho_{0}$ which is given in ${\rm fm}^{-3}$.}
\label{Table2}
\end{table}
\end{center}
We observe from Table\,\ref{Table2} that, with the exception of $L$, all three models are in very close agreement. It is important to note that the calibration of FSUGarnet relied exclusively on physical observables that can be either measured in the laboratory or extracted from observation, so the bulk properties listed in Table\,\ref{Table2} are genuine model predictions. 
\begin{figure}[ht]
 \centering
 \includegraphics[width=0.4\textwidth]{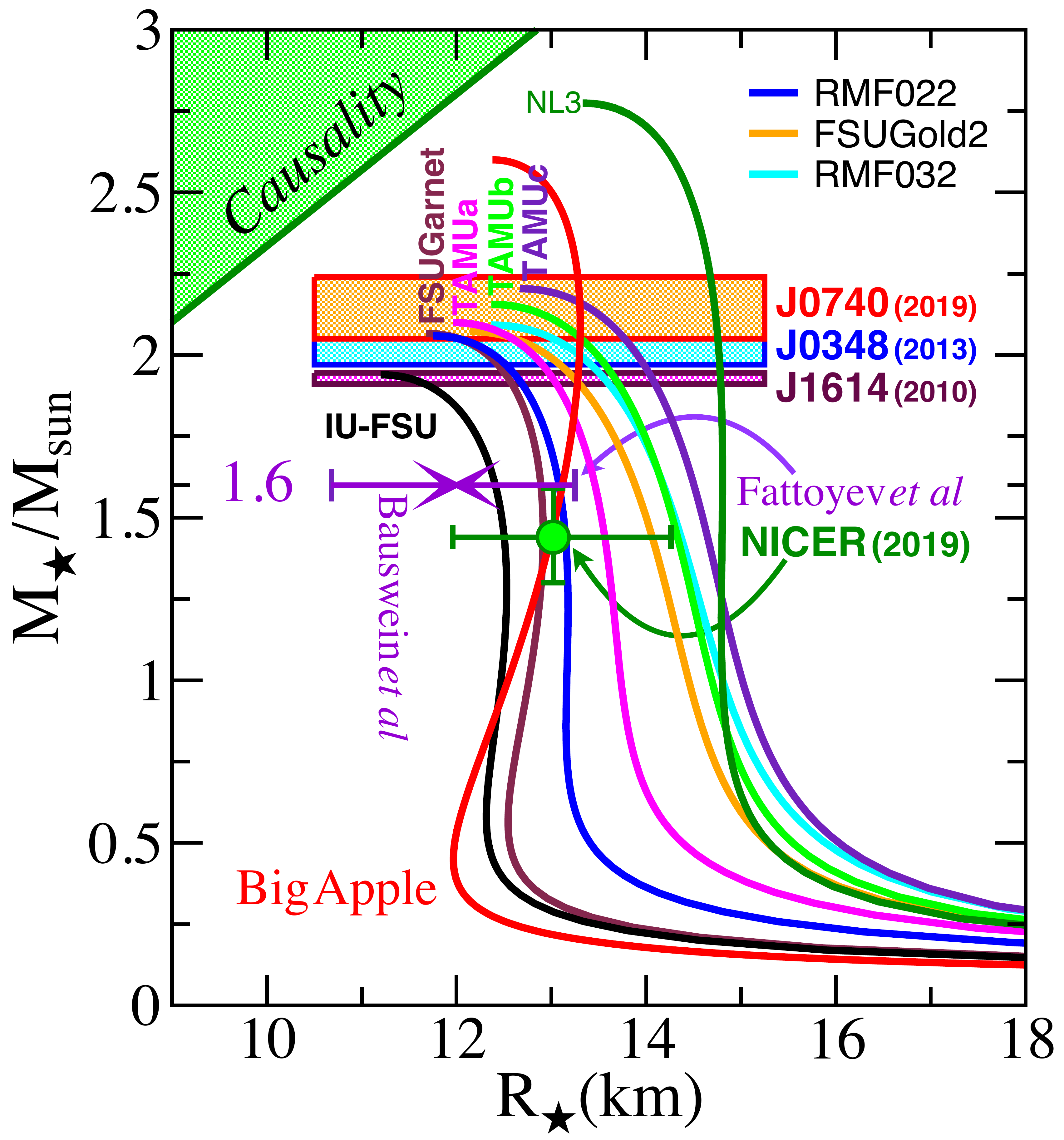}
 \caption{(Color online). Mass-vs-Radius relation predicted by a variety of covariant EDFs. The figure has been adapted from Ref.\,\cite{Fattoyev:2017jql} and is supplemented by the maximum mass constraint from Ref.\,\cite{Cromartie:2019kug}, NICER constraint on the stellar radius of a 1.4\,$M_{\odot}$ neutron star\,\cite{Riley:2019yda,Miller:2019cac}, and our predictions from BigApple.}
\label{Fig1}
\end{figure}

Displayed  in Fig.\,\ref{Fig1} is the mass-{\sl vs}-radius relation as predicted by a variety of covariant EDFs---including BigApple. The figure has been adapted from Ref.\,\cite{Fattoyev:2017jql} and includes, besides earlier constraints on the maximum neutron star mass\,\cite{Demorest:2010bx, Antoniadis:2013pzd}, the newest constraint from Cromartie et al.\,\cite{Cromartie:2019kug} as well as the recent NICER result. All models support a $\sim\!2\,M_{\odot}$ neutron star and, with couple of exceptions, agree with the NICER result. Yet, several of these models generate stellar radii for medium mass neutron stars that are inconsistent with the tidal deformability extracted from GW170817\,\cite{Abbott:PRL2017}. Note, however, that BigApple---with a maximum neutron-star mass of 2.6\,$M_{\odot}$ satisfies all observational constraints; see also Table\,\ref{Table3}.

\begin{center}
\begin{table}[h]
\begin{tabular}{| l | | c | c | c || c | c |}
 \hline\rule{0pt}{2.5ex} 
\!\!Model  &  $R_{1.4}$     &  $\Lambda_{1.4}$ & $M_{\rm max}$ &  
                    $R_{\rm max}$ & $\Lambda_{\rm max}$ \\  
\hline
 \hline\rule{0pt}{2.5ex} 
\!\!IUFSU    & 12.528 &   499.2   &   1.939    &   11.265   &  20.9 \\
FSUGarnet    & 12.869 &   624.8   &   2.066    &   11.706  &  18.2  \\
BigApple     & 12.960 &   717.3   &   2.600    &   12.417   &  5.0  \\
\hline
\end{tabular}
\caption{Stellar properties as predicted by the three covariant EDFs used in this work. The maximum mass is given in solar masses, both radii in km, and the tidal deformability is dimensionless.}
\label{Table3}
\end{table}
\end{center}

\begin{figure}[ht]
 \centering
 \includegraphics[width=0.45\textwidth]{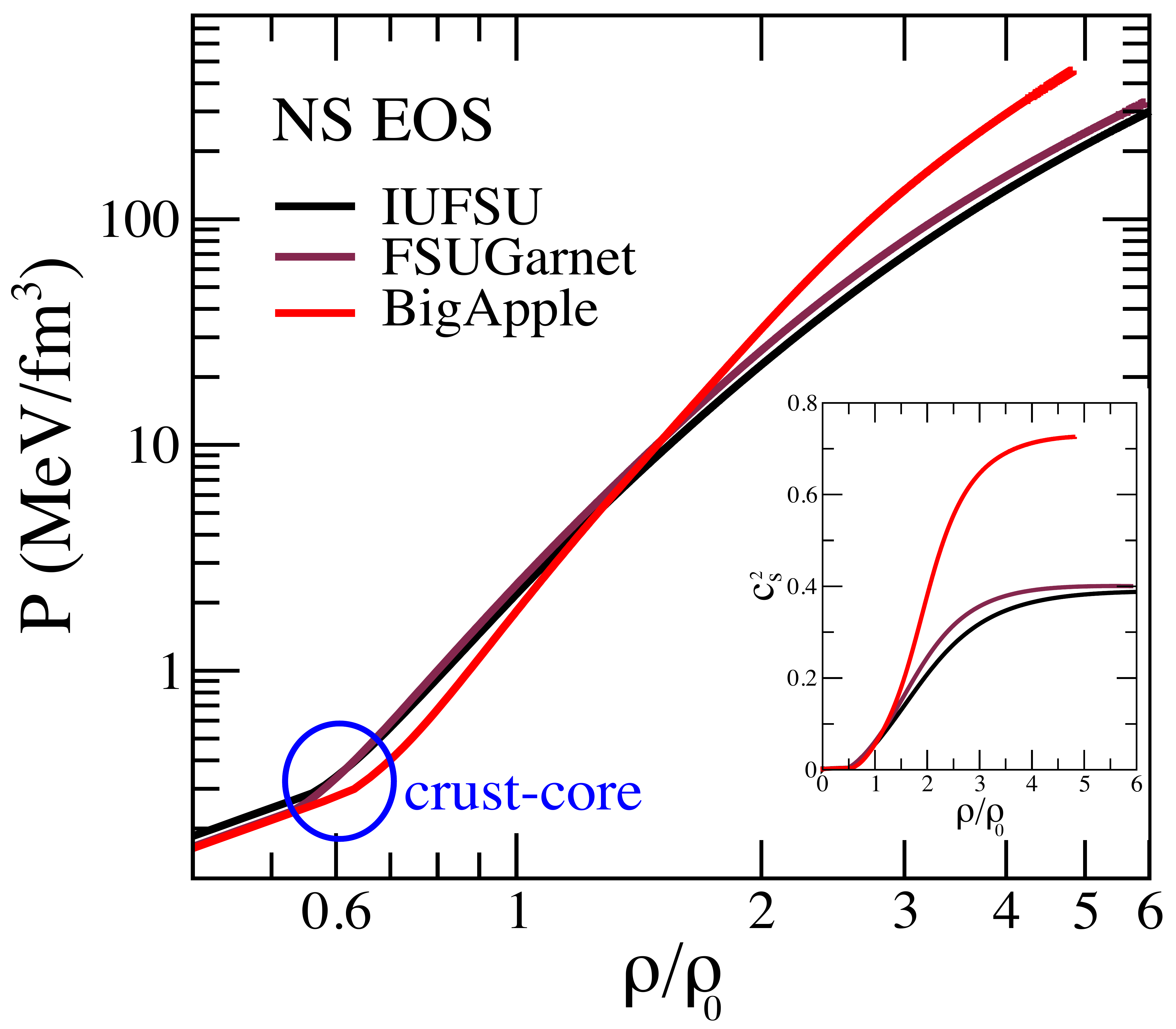}
 \caption{(Color online) Equation of state---pressure as a function of baryon density---for neutron-star matter in chemical equilibrium. Here $\rho_{0}\!\simeq\!0.154\,{\rm fm}^{-3}$ is the density of nuclear matter at saturation (see Table\,\ref{Table2}), and the inset displays the associated speed of sound in units of 
 the speed of light.}
\label{Fig2}
\end{figure}

We now proceed to examine the microphysics responsible for the macroscopic properties displayed in Fig.\,\ref{Fig1}. The underlying neutron-star matter EOS alongside the square of the speed of sound---defined as the derivative of the pressure with respect to the energy density---is displayed in Fig.\,\ref{Fig2}. The kink in the pressure at around $2/3$ saturation density signals the transition from the solid crust to the liquid core. At saturation density ($\rho_{0}$) the pressure is generated exclusively by the symmetry energy. This pressure determines the thickness of the neutron skin in neutron-rich nuclei\,\cite{Brown:2000, Furnstahl:2001un, Centelles:2008vu, RocaMaza:2011pm}. In turn, the pressure at two-to-three times saturation density appears to controls the radius of medium-mass neutron stars\,\cite{Lattimer:2006xb, Tews:2019cap, Capano:2019eae, Drischler:2020yad}. Indeed, all three models share the same pressure just below 2$\rho_{0}$ and, as consequence, predict similar radii for a canonical 1.4\,$M_{\odot}$ neutron star. However, the maximum mass of stable neutron stars stars is highly sensitive to the pressure at the highest densities\,\cite{Chatziioannou:2020pqz}. Indeed, a dramatic rate of increase in the pressure---best reflected in the speed of sound---is required to support a super-massive 2.6\,$M_{\odot}$ neutron star.

Based on the evidence presented so far, there seems to be no compelling argument against the possible existence of a super-massive 2.6\,$M_{\odot}$ neutron star. Note, however, that whereas the relatively large tidal deformability predicted by BigApple is consistent with the limits presented in the discovery paper\,\cite{Abbott:PRL2017}, the revised limit of $\Lambda_{1.4}\!=\!190^{+390}_{-120}$\,\cite{Abbott:2018exr} presents a more serious challenge. However, the most serious evidence against such a stiff EOS comes from laboratory experiments involving the energetic collision between two gold nuclei, a violent encounter that compressed matter to pressures in excess of $10^{34}$\,Pa\,\cite{Danielewicz:2002pu}; 
note that 1Pa\,=10 dyn/cm$^{2}$=\,6.242$\times10^{-33}{\rm MeV/fm}^{3}$. 

\begin{figure}[ht]
 \centering
 \includegraphics[width=0.45\textwidth]{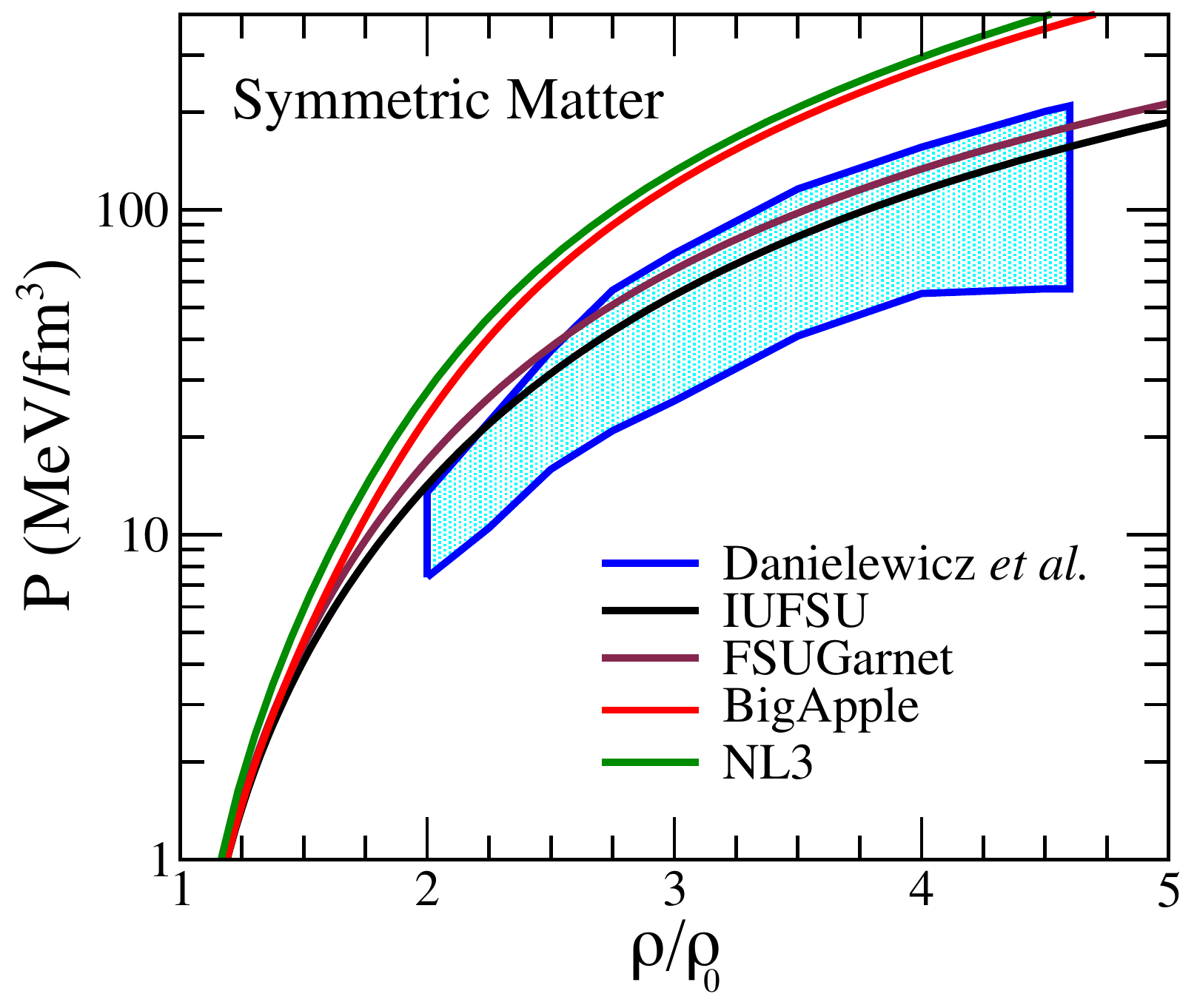}
 \caption{(Color online) Equation of state---pressure as a function of baryon density---of symmetric nuclear matter. The blue shaded area represents the EOS extracted from the analysis of flow data using a value of $\rho_{0}\!\simeq\!0.16\,{\rm fm}^{-3}$ for the saturation density\,\cite{Danielewicz:2002pu}.}
\label{Fig3}
\end{figure}

We display in Fig.\,\ref{Fig3} constraints on the EOS of symmetric nuclear matter as extracted from the analysis of particle flow in heavy-ion collisions\,\cite{Danielewicz:2002pu}. We observe that IUFSU and FSUGarnet---with predictions of 1.94\,$M_{\odot}$ and 2.07\,$M_{\odot}$ for the maximum neutron star mass---already sit near the upper edge of the allowed region. In contrast, the very stiff EOS predicted by the NL3 parametrization\,\cite{Lalazissis:1996rd} was explicitly ruled out by the heavy-ion data\,\cite{Danielewicz:2002pu}. Following a similar trend as NL3, it is clear that the stiff EOS predicted by BigApple and required to account for super-massive neutron stars is also ruled out. Given that FSUGarnet is approaching the upper boundary allowed by the heavy-ion data, it is unlikely that the maximum neutron star mass can go much beyond the present observational limit of 2.14\,$M_{\odot}$\,\cite{Cromartie:2019kug}. 

In summary, motivated by the recent identification of a compact object in the $2.5\!-\!5.0\,M_{\odot}$ mass gap\,\cite{Abbott:2020khf}, we have explored the possibility that such an object could be a super-massive neutron star. Given the lack of an electromagnetic counterpart to GW190814 and the absence of tidal distortions to the gravitational waveform, it is unlikely that the nature of the 2.6\,$M_{\odot}$ compact object will ever be resolved by a further analysis of the data. In this letter we have adapted a class of modern covariant EDFs to account for a stable 2.6\,$M_{\odot}$ neutron star, while ensuring that earlier constraints on the structure of neutron stars as well as ground-state properties of finite nuclei are accurately reproduced. Indeed, we demonstrated that such an EDF---dubbed ``BigApple"---successfully accounts for the binding energy and charge radii of a representative set of spherical nuclei, is consistent with the bulk properties of infinite nuclear matter, reproduces the recent NICER data, and is compatible with the limits on the tidal deformability of a 1.4\,$M_{\odot}$ neutron star, as reported in the GW170817 discovery paper. In particular, our predictions are well aligned (see Table\,\ref{Table3}) with the constraints extracted from GW190814 under the NSBH scenario: $R_{1.4}\!=\!12.9^{+0.8}_{-0.7}$ and $\Lambda_{1.4}\!=\!616^{+273}_{-158}$\,\cite{Abbott:2020khf}.

However, despite the considerable success of the model, two sets of data strongly disfavor such a stiff EOS. First, the revised upper limit from GW170817 of $\Lambda_{1.4}\!=\!190^{+390}_{-120}$ (at the 90\% level)\,\cite{Abbott:2018exr} is significantly lower than the $\Lambda_{1.4}\!\simeq\!720$ value predicted by BigApple. This would require a softening of the EOS, particularly of the symmetry energy. Second, constraints on the EOS of symmetric nuclear matter extracted from particle-flow in high-energy nuclear collisions rule out an overly stiff nuclear matter EOS. For example, BigApple predicts a pressure at four times saturation density that is nearly twice as large as the upper limit extracted from the flow data. In principle, by adding additional interactions at high densities one could soften the EOS of symmetric nuclear matter to bring it into agreement with the heavy-ion data at the expense of needing to stiffen the symmetry energy. Whereas such a procedure may still result in an overall EOS that can account for a 2.6\,$M_{\odot}$ neutron star, it may require considerable fine tuning to keep the pressure of nuclear matter and the tidal deformability low enough.  Hence, we conclude that the low deformability demanded by GW170817 combined with the heavy-ion data for symmetric nuclear matter make it highly unlikely that the maximum mass could be as large as 2.6\,$M_{\odot}$, at least for the class of models used in this work. So as one is left speculating whether the 2.6\,$M_{\odot}$ compact object in GW190814 is either the most massive neutron star or the lightest black hole ever detected, our analysis points strongly in favor of the latter.
\begin{acknowledgments}\vspace{-10pt}
 This material is based upon work supported by the U.S. Department of Energy Office of Science, Office of Nuclear Physics under Awards DE-FG02-87ER40365 (Indiana University), Number DE-FG02-92ER40750 (Florida State University), and Number DE-SC0008808 (NUCLEI SciDAC Collaboration).
\end{acknowledgments} 
\vfill\eject

\bibliography{GW190814.bbl}

\vfill\eject
\end{document}